\documentclass[onecollarge,natbib]{svjour2}
\bibpunct{[}{]}{;}{n}{}{,} 
\smartqed  
\usepackage{graphicx}
\usepackage{latexsym}
\newcommand{\bm}[1]{\mbox{\boldmath $#1$}}

\newcommand{\open}{{<\kern -0.3 em{\scriptscriptstyle )}}}

\newcommand{\nslash}{\kern 0.2 em n\kern -0.45em /}
\newcommand{\Pslash}{\kern 0.2 em P\kern -0.56em \raisebox{0.3ex}{/}}
\newcommand{\pslash}{\kern 0.2 em p\kern -0.4em /}
\newcommand{\kslash}{\kern 0.2 em k\kern -0.45em /}
\newcommand{\Sslash}{\kern 0.2 em S\kern -0.56em \raisebox{0.3ex}{/}}

\newcommand{\eq}{\begin{equation}}
\newcommand{\ee}{\end{equation}}
\newcommand{\beq}{\begin{equation}}
\newcommand{\eeq}{\end{equation}}
\newcommand{\ba}{\begin{eqnarray}}
\newcommand{\ea}{\end{eqnarray}}

\newcommand{\sumint}{\kern 0.2 em {\textstyle\sum} \kern -1.1 em \int}

\journalname{Few-Body Systems}
\begin{document}

\title{Gluon TMDs in quarkonium production} 

\author{Dani\"el Boer}


\institute{D. Boer \at 
Van Swinderen Institute for Particle Physics and Gravity\\
University of Groningen, The Netherlands\\
                \email{d.boer@rug.nl} }

\date{November 18, 2016}

\maketitle

\begin{abstract}
Quarkonium production offers good possibilities to study gluon TMDs. In this proceedings contribution this topic is explored for the linearly polarized gluons inside unpolarized hadrons and unpolarized gluons inside transversely polarized hadrons. It is argued that $\chi_{b0/2}$ and $\eta_b$ production at LHC are best to study the effects of linearly polarized gluons in hadronic collisions, by means of angular independent ratios of ratios of cross sections. This can be directly compared to $\cos 2\phi$ asymmetries in heavy quark pair and dijet production in DIS at a future high-energy Electron-Ion Collider (EIC), which probe the same TMDs. In the small-$x$ limit this corresponds to the Weizs\"acker-Williams (WW) gluon distributions, which should show a change in behavior for transverse momenta around the saturation scale. Together with investigations of the dipole (DP) gluon distributions, this can provide valuable information about the polarization of the Color Glass Condensate if sufficiently small $x$ are reached. Quarkonia can also be useful in the study of single transverse spin asymmetries. For transversely polarized hadrons the gluon distribution can be asymmetric, which is referred to as the Sivers effect. It leads to single spin asymmetries in for instance $J/\psi$ (pair) production at AFTER@LHC, which probe the WW or $f$-type gluon Sivers TMD. It allows for a test of a sign-change relation w.r.t.\ the gluon Sivers TMD probed at an EIC in open heavy quark pair production. Single spin asymmetries in backward inclusive $C$-odd quarkonium production, such as $J/\psi$ production, may offer probes of the DP or $d$-type gluon Sivers TMD at small $x$-values in the polarized proton, which in that limit corresponds to a correlator of a single Wilson loop, describing the spin-dependent odderon.  
\end{abstract}

\section{Introduction}
Inclusive quarkonium production processes offer good possibilities to probe gluons inside protons.
Especially in case of transverse momentum dependent observables they offer the opportunity to access the largely unknown 
transverse momentum dependent gluon distribution functions, the so-called gluon TMDs. Because of the additional dependence 
on the transverse momentum vector, there are more gluon TMDs than collinear gluon densities: at leading twist there are eight 
gluon TMDs, compared to only two collinear gluon pdfs ($g(x)$ and $\Delta g(x)$). 
This means that six gluon TMD structures average to zero, when integrated 
over all transverse momenta. One of these describes the gluon Sivers effect \cite{Sivers:1989cc}, which is odd in the transverse momentum. Another describes the distribution of linearly polarized gluons, that is even in the transverse momentum, but corresponds to a $\cos 2\phi$ dependence that 
averages away too. These functions appear in the parametrization of the hadronic gluon correlator $\Gamma$ \cite{Mulders:2000sh}
\beq
\Gamma^{\,\mu\nu\, [{\cal U},{\cal U}^\prime]}(x,\bm{k}_T) \equiv \left. \int \frac{d(\xi\cdot P)\, d^2 \xi_T}{(P\cdot n)^2 (2\pi)^3} e^{i ( xP + k_T) \cdot \xi} \langle P| {\rm Tr}_c \left[ F^{n\nu}(0)\, {\cal U}_{[0,\xi]}\, F^{n\mu}(\xi)\, {\cal U}_{[\xi,0]}^\prime \right] |P\rangle\right|_{\xi \cdot n = 0}. \label{GammaUU} 
\eeq
Although this correlator is parametrized in terms of eight leading twist gluon TMDs, in analogy to quark TMDs, in this proceedings contribution only three will be discussed in detail: the unpolarized gluon distribution function, the linearly polarized gluon distribution and the gluon Sivers function. 
The process dependence that enters through the gauge links, ${\cal U}$ and ${\cal U}^\prime$, plays an important role. 
It implies that for each of the TMDs there are (at least) two versions to consider, which are independent of each other and that generally behave differently as a function of $x$ and $k_T^2$. This will be discussed below, in particular when we consider the limit of small $x$-values. The ${\cal U}$ and ${\cal U}^\prime$ dependence will sometimes be suppressed in what follows, but is understood to be present for all TMDs. 

For unpolarized hadrons the gluon correlator $\Gamma_{U}$ is parametrized by two TMDs (here $\bm{k}_T^2= -k_T^2$):
\beq
\Gamma_{U}^{\,\mu\nu}(x,\bm{k}_T ) = \frac{x}{2}\,\bigg \{-g_T^{\mu\nu}\,f_1^{\, g}(x,\bm{k}_T^2)
+\bigg(\frac{k_T^\mu k_T^\nu}{M_p^2}\, {+}\,g_T^{\mu\nu}\frac{\boldmath{k}_T^2}{2M_p^2}\bigg)
\;h_1^{\perp\,g}(x,\bm{k}_T^2) \bigg \}, 
\eeq
where $f_1^{\, g}$ is the unpolarized gluon TMD and $h_1^{\perp\, g}$ the linearly polarized gluon TMD.  
Even the unpolarized gluon TMD has not been extracted from experiments yet. According to~\cite{Dunnen:2014eta} the processes of $\Upsilon + \gamma$ and $J/\psi + \gamma$ production at the LHC offer good possibilities for this, where in the $\Upsilon$ case the color singlet contribution dominates and in the $J/\psi$ case also for not too large invariant mass of the $J/\psi + \gamma$ system.
The linear gluon polarization can be accessed through azimuthal modulations in these processes too, or through the transverse momentum squared distribution of Higgs (H) production \cite{Sun:2011iw,Boer:2011kf} and of heavy $C$-even (pseudo-)scalar quarkonium production \cite{Boer:2014tka,Echevarria:2015uaa,Signori:2016jwo}.
The latter process will be denoted here as $p\,p \to [\overline{Q}Q]\, X$, where $[\overline{Q}Q]=\chi_{c0},\eta_c,\chi_{b0},\eta_b$. TMD evolution studies indicate that the lower the mass of the quarkonium state, the larger the effects from linear gluon polarization, but also the larger the uncertainties from nonperturbative QCD contributions \cite{Boer:2015uqa,Echevarria:2015uaa}. In this respect bottomonium states seem to offer the best balance. 

In general, the hadron production process $p \,  p \to h\, X$, where $h$ denotes a light hadron, is not a TMD process. Factorization requires large transverse momentum of the hadron $h$, leading to collinear factorization. The situation is different when the hadron is heavy, like for a quarkonium state, because its large mass provides an additional large scale in which to expand. In the case of a $C$-even quarkonium state, the production mechanism is primarily that of gluon-gluon fusion, like in Higgs production. If the quarkonium state is produced from two gluons in a color singlet state, then the process $p\,p \to [\overline{Q}Q]\, X$ is similar to Drell-Yan or Higgs production and can be treated as a TMD process. 
Two gluons in the color singlet state will only produce a $C$-even $Q\overline{Q}$ state~\cite{Novikov:1977dq}, hence $C$-even quarkonia of $J\neq 1$ can be in the color singlet state when produced from the fusion of two gluons\footnote{The $J=1$ case is excluded because of the Landau-Yang theorem, which applies only to the color singlet case \cite{Beenakker:2015mra}. $C$-odd quarkonia can only be in the color octet state when produced from two gluons. The $C$-odd $J/\psi$ or $\Upsilon$ production requires the presence of three gluons to be in the color singlet state and will involve the color octet state in general.}. $C$-even quarkonia can also be in the color octet state, but NRQCD can be used to argue that those contributions are suppressed in this case. For $C=+$ {\it bottomonium} production the color singlet contribution dominates, not only according to NRQCD considerations \cite{Hagler:2000dd,Bodwin:2005hm} but also according to several numerical studies \cite{Hagler:2000dd,Maltoni:2004hv,Lansberg:2012kf}. This forms the motivation for using the color singlet model \cite{Baier:1983va} for $C$-even quarkonium production. 

Using the color singlet model and leading order (LO) NRQCD \cite{Bodwin:1994jh,Bodwin:2005hm}, the differential cross sections for $\eta_b$, $\chi_{b 0}$ and $\chi_{b 2}$ production can be obtained \cite{Boer:2012bt}, which allow to probe the relative contribution from linearly polarized gluons described by the ratio ${\cal R}(Q_T)$~\cite{Boer:2011kf}:
\beq
{\cal R}(Q_T) \equiv \frac{\mathcal{C}[w_H\,h_1^{\perp\, g}\,h_1^{\perp\, g}]}{\mathcal{C}[f^{\, g}_1\,f_1^{\, g}]} =  \frac{\int d^2 \bm{b} \, e^{i \bm{b} \cdot \bm{q}_T^{}} e^{-S_A(b_*,Q) - S_{NP}(b,Q)} \; \widetilde{h}_1^{\perp\, g}(x_{A},b_*^2; \mu_{b_*}) \; \widetilde{h}_1^{\perp\, g}(x_{B},b_*^2; \mu_{b_*})}{
\int d^2 \bm{b} \, e^{i \bm{b} \cdot \bm{q}_T^{}} \, e^{-S_A(b_*,Q)- S_{NP}(b,Q)} \widetilde{f}_1^{\, g}(x_{A},b_*^2; \mu_{b_*}) \, 
\widetilde{f}_1^{\, g}(x_{B},b_*^2; \mu_{b_*})}. 
\label{calRQT}
\eeq
Here $Q_T^2= \bm{q}_T^2$, ${\cal C}$ denotes a convolution of TMDs, $w_{H}=\left((\bm{k}_{1T}\cdot\bm{k}_{2T})^{2}-\frac{1}{2}\bm{k}_{1T}^{2} \bm{k}_{2T}^{2}\right)/2M^4$, $\widetilde{f}_1^{\, g}$ denotes the Fourier transform of $f_1^{\, g}$, and similarly for $h_1^{\perp\, g}$:
\beq
\widetilde{h}_1^{\perp\,g}(x,b^2) \equiv \int d^2\bm{k}_T^{}\; \frac{(\bm{b}\!\cdot \! \bm{k}_T^{})^2 - \frac{1}{2}\bm{b}^{2} \bm{k}_T^{2}}{b^2 M^2}
\; e^{-i \bm{b} \cdot \bm{k}_T^{}}\; h_1^{\perp\, g}(x,k_T^2) =  -\pi \int dk_T^2 \frac{k_T^2}{2M^2} J_2(bk_T) h_1^{\perp\, g}(x,k_T^2).
\eeq
For details about the Sudakov factors $S_A$ and $S_{NP}$, $\mu_{b_*}$ and $b_*$ prescription, see \cite{Boer:2015uqa}. 

By forming ratios of ratios of cross sections, hadronic uncertainties drop out, leading to \cite{Boer:2015uqa}: 
\beq
\frac{\sigma (\chi_{b 2})}{\sigma(\chi_{b 0})} \frac{d\sigma(\chi_{b 0})/d^2 \bm{q}_T}{d\sigma (\chi_{b 2})/d^2 \bm{q}_T}  \approx  1 + {\cal R}(Q_T), \qquad
\frac{\sigma (\chi_{b 0})}{\sigma(\eta_b)} \frac{d\sigma(\eta_b)/d^2 \bm{q}_T}{d\sigma (\chi_{b 0})/d^2 \bm{q}_T} \approx  \frac{1 - {\cal R}(Q_T)}{1 + {\cal R}(Q_T)}.\label{ratioofratios}
\eeq
It should be mentioned that the TMD factorization for the $p$-wave states $\chi_{bJ}$ has been called into question in \cite{Ma:2014oha}. However, the comparison of several of these ratios of ratios, {\it including} the $p$-wave states $\chi_{bJ}$, allows to test for such TMD factorization breaking effects and in case these effects turn out to be small, they allow for an extraction of ${\cal R}(Q_T)$. Another advantage of this comparison is that due to the small energy scale differences between the three bottomonia, i.e.\ $m_{\eta_b} =9.4$ GeV, $m_{\chi_{b0}}=9.9$ GeV, and $m_{\chi_{b2}} =10.3$ GeV, evolution effects should play only a negligible role. A drawback of this suggestion is that the $J=0$ bottomonia are quite challenging to measure experimentally at small $Q_T$. 

For transversely polarized hadrons the spin ($S_T$) dependent part $\Gamma_T$ of the gluon correlator is parametrized by four TMDs (one T-even and three T-odd), where we only display the T-odd Sivers term:
\beq
\Gamma_{T}^{\,\mu\nu}(x,\bm{k}_T )
= \frac{x}{2}\,\bigg \{g^{\mu\nu}_T\,    \frac{ \epsilon^{\rho\sigma}_T k_{T \rho}\, S_{T\sigma}}{M_p}\, f_{1T}^{\perp\,g}(x,\bm{k}_T^2) + \ldots \bigg \}, 
\eeq
The T-odd functions appear in the description of single transverse spin asymmetries. The current knowledge on the gluon Sivers TMD is reviewed in \cite{Boer:2015vso}, where it is  discussed that the most promising processes to probe it are: $p^\uparrow\,p \to \gamma \,{\rm jet}\, X$ \cite{Schmidt:2005gv,Bacchetta:2007sz} at RHIC and at a polarized fixed-target experiment at LHC (AFTER@LHC), $p^\uparrow\,p \to J/\psi\, \gamma \, X$ at AFTER@LHC \cite{Lansberg:2014myg}, and $e \, p^\uparrow \to e'\, Q \, \overline{Q} \, X$ at a possible future Electron-Ion Collider \cite{Boer:2011fh}. For these processes TMD factorization needs to be proven (or disproven) still, cf.\ e.g.\ \cite{Zhu:2013yxa}, which provides a caveat for the discussion below. Proceeding under the assumption of TMD factorization, we need to consider the dependence on the gauge links, ${\cal U}$ and ${\cal U}^\prime$, mentioned above. It turns out that $p^\uparrow\,p \to \gamma \,{\rm jet}\, X$ probes a different Sivers function than $p^\uparrow\,p \to J/\psi\, \gamma \, X$ and $e \, p^\uparrow \to e'\, Q \, \overline{Q} \, X$.

\section{Process dependence} 
From studies of single spin asymmetries it has become clear that TMDs in general are not universal \cite{Brodsky:2002cx,Collins:2002kn,Belitsky:2002sm}. The process dependent gauge links ${\cal U}$ and ${\cal U}^\prime$ in the gluon correlator arise from summing gluon rescattering corrections \cite{Efremov:1978xm}. In TMDs the gauge links are path-ordered exponentials ${\cal U}_{{\cal C}}$ that depend on a path ${\cal C}$ that is not entirely along the lightcone:
\beq
{\cal U}_{{\cal C}}[0,\xi] = {\cal P} \exp \left(-ig\int_{{{\cal C}[0,\xi]}} ds_\mu \, A^\mu (s)\right),
\eeq
where $\xi=[0^+,\xi^-,\bm{\xi}_T]$. The path depends on the process, i.e.\ on whether color charges are incoming and/or outgoing. This was recognized a long time ago \cite{Collins:1983pk,Boer:1999si}, but it was generally thought that gauge links were irrelevant because a gauge invariant quantity should yield the same result in any gauge and the gauge link becomes unity in a specific gauge, hence removing the path dependence from the operator in that gauge. But the fields then depend on that gauge, and as such they now become dependent on the path. {\it A priori} it is not clear that this path dependence will affect observables, but it turns out that it does. This became clear from the study of single spin asymmetries \cite{Collins:2002kn,Brodsky:2002rv,Belitsky:2002sm,Boer:2003cm} arising from the Sivers effect. By now it is well-known that in semi-inclusive DIS the quark TMD correlator contains a future pointing staple-like Wilson line (referred to as a $+$ link) arising from final state interactions (FSI). In the Drell-Yan (DY) process the path is past pointing (referred to as a $-$ link) arising from initial state interactions (ISI). The quark Sivers TMDs with $+$ and $-$ links are related by parity and time reversal invariance, yielding the TMD formalism prediction $f_{1T}^{\perp [{\rm SIDIS}]} = - f_{1T}^{\perp [{\rm DY}]}$ \cite{Collins:2002kn}. A similar relation holds for gluon Sivers TMDs \cite{Boer:2016fqd}: 
\begin{equation}
f_{1T}^{\perp\, g \, [e\, p^\uparrow \to e' \, Q\, \overline{Q}\, X]}(x,p_T^2) = - f_{1T}^{\perp\, g \, [p^\uparrow\,  p\to \gamma \, \gamma \, X]} (x,p_T^2),
\end{equation}
where the transverse momentum of the produced pair of heavy quarks and photons is much smaller than the invariant mass of the pair. 
The gluon rescatterings in the subprocess $\gamma^*\, g \to Q\, \overline{Q}$ lead to two future pointing Wilson lines in the gluon TMD correlator, whereas the subprocess $g \, g \to \gamma \gamma$ (which is dominant in $p\,  p\to \gamma \, \gamma \, X$ in the back-to-back correlation limit and central photon pair rapidity \cite{Qiu:2011ai}) leads to two past-pointing Wilson lines. The above relation thus reflects that $f_{1T}^{\perp \, g \, [+,+]}= - f_{1T}^{\perp \, g \, [-,-]}$. Both TMD formalism sign change predictions need to be verified in experiments still. Instead of a photon pair one can consider other color singlet states in
$gg$-dominated kinematics, like for instance $J/\psi \,\gamma$ or $J/\psi\, J/\psi$~\cite{Dunnen:2014eta,Lansberg:2014myg,Lansberg:2015lva}. These processes can hopefully be studied at RHIC or AFTER@LHC and compared to EIC data in the future. 

As the gluon TMDs depend on two gauge links, one can also consider $f_{1T}^{\perp \, g \, [+,-]}= - f_{1T}^{\perp \, g \, [-,+]}$. The process
$p^\uparrow\,p \to \gamma \,{\rm jet}\, X$ probes $f_{1T}^{\perp \, g \, [+,-]}$ in the kinematic region where the subprocess $q \, g \to \gamma \, q$ dominates. Hence, this process probes an {\it a priori} completely independent gluon Sivers function. The first transverse moment of $f_{1T}^{\perp \, g \, [+,+]}$ and $f_{1T}^{\perp \, g \, [-,+]}$ involve antisymmetric ($f_{abc}$) and symmetric ($d_{abc}$) color structures, respectively, as discussed in e.g.\ \cite{Buffing:2013kca}. Therefore, these two Sivers functions are also sometimes referred to as $f$-type and $d$-type functions. We note that this terminology only makes sense for the T-odd TMDs, whose transverse moments are related to triple gluon matrix elements. 

The more hadrons observed in a process, the more complicated the color charge flow. The resulting Wilson lines can be combinations of $+$ and $-$ links, possibly with additional loops, leading to more complicated relations among TMDs of various processes. The various Wilson lines can even become entangled, such that the color traces cannot be disentangled, leading to factorization breaking contributions \cite{Collins:2007nk,Rogers:2010dm}. They may perhaps be disentangled for certain transverse moments \cite{Bomhof:2004aw,Buffing:2012sz,Buffing:2013dxa}. This entanglement does not only jeopardize predictions, it may also lead to additional angular dependences that are otherwise absent \cite{Rogers:2013zha}. At present nothing is known about the size of this type of factorization breaking contributions. It is expected to affect $p \, p \to h_1 \, h_2 \, X$ processes, where the two hadrons in the final state can also be quarkonium states, such as the $J/\psi$ pair case discussed above. However, quarkonia do introduce one or more additional large scales into the problem, which might be used to establish factorization for those cases. This remains to be studied. It should be mentioned that the TMD factorization breaking for the $p$-wave states $\chi_{bJ}$ discussed in \cite{Ma:2014oha} is of a different kind. 

The TMD nonuniversality is not only observable in scattering with polarized protons. Also in unpolarized scattering one has to deal with multiple gluon distributions (as first became clear in small-$x$ physics \cite{Dominguez:2011wm}, see below). This process dependence of unpolarized TMDs also implies that the $p_T$-widths of these distributions are process dependent. It gives an 
additional process dependence to the $p_T$-broadening observable. However, one can show that each width (i.e.\ the average $p_T^2$) can be expanded in terms of five independent universal widths with calculable integer coefficients \cite{Boer:2015kxa}, which limits the possibilities somewhat. 

\section{Small-$x$ limit}
Quarkonium production is often considered because of its sensitivity to gluons. Similarly, the small-$x$ limit is considered for this purpose, since the gluons dominate at small $x$. The combination of quarkonium production at small $x$ may thus be expected to yield rather clean signals of gluon TMD effects, because the small $x$-values enhance the gluon contributions and the quarkonium acts as a filter, reducing the contribution from quarks even further. The small-$x$ limit also allows consideration of a larger class of processes, because  certain factorization breaking partonic subprocesses may become suppressed in this limit, resulting effectively in a restoration of TMD factorization or in some form of hybrid factorization involving both TMDs and collinear distributions \cite{Chirilli:2011km,Kotko:2015ura}. The latter is often considered for the process $p A \to h\, X$. In this section we will look at what are the possibilities to study the gluon TMDs at small $x$, including several quarkonium production processes. 

\subsection{Unpolarized gluon TMDs at small $x$}
For most processes of interest there are two unpolarized gluon distributions to consider: $G^{(1)} \equiv f_1^{\,g \, [+,+]}= f_1^{\,g \, [-,-]}$ and $G^{(2)}\equiv f_1^{\,g \, [+,-]}= f_1^{\,g \, [-,+]}$~\cite{Dominguez:2011wm}. At small $x$ these two distributions are referred to as the Weizs\"acker-Williams (WW) and dipole (DP) gluon distributions. Kharzeev, Kovchegov, and Tuchin (KKT) were the first to observe that there are two distinct but equally valid definitions for the gluon distribution at small $x$~\cite{Kharzeev:2003wz}. They stated: ``The authors cannot offer any simple physical
explanation of this paradox.'' It is now understood that the difference comes from which process is considered. The WW and DP distributions would be equal without any ISI or FSI (and unlike the Sivers TMD they would still be nonzero in that case). Moreover, KKT employed the McLerran-Venugopalan (MV) model in which case the two gluon distributions can be related \cite{Kharzeev:2003wz, Dominguez:2011wm}. For example, in the MV model $\gamma \, A \to Q \, \overline{Q} \, X$ can be expressed in terms of the DP distribution $C(k_\perp)= \int d^2x_\perp e^{ik_\perp \cdot x_\perp} \langle U(0)U^\dagger(x_\perp)\rangle \sim G^{(2)}$ \cite{Gelis:2001da}, although it probes the WW distribution. A natural question that arises is how different these two distributions can be? The only constraints are that the integrated distributions ($\int d^2 \bm{k}_T$) have to be the same (model independently) and the functions need to match onto the same universal perturbative large-$k_T$ tail. This allows for different shapes and magnitudes. Therefore, it is worthwhile to study experimentally different processes that probe either $G^{(1)}$ or $G^{(2)}$. Several processes are listed in Table \ref{f1table}, which is in part based on \cite{Dominguez:2011wm}.
\begin{table}[htb]
\centering
\caption{Selection of processes that probe the WW or DP unpolarized gluon TMD at small $x$.}
\label{f1table}       
\begin{tabular}{|l|c|c|c|c|c|c|c|}
\hline
\hline
{}& {} & {} & {} & {} & {} &{} & {}\\[-2 mm]
{}& DIS & DY & SIDIS & $pA\to \gamma\, {\rm jet}\, X$ & $e \, p \to e'\, Q \, \overline{Q} \, X$ & $p p \to \eta_{c,b} \, X$ & $pp \to J/\psi\, \gamma\, X$ \\[0.4 mm]
{}& {} & {} & {} & {} & $e \, p \to e'\, j_1 \, j_2 \, X$ & $p p \to H \, X \hspace{2mm} $ & $pp \to \Upsilon\, \gamma\, X\hspace{3mm} $ \\[0.4mm]\hline
$f_1^{\,g \, [+,+]}$ (WW) & $\times$ &$\times$ & $\times$ & $\times$ & $\surd$ & $\surd$ & $\surd$ \\[0.6 mm]\hline
$ f_1^{\,g\, [+,-]}$ (DP) & $\surd$ &$\surd$ & $\surd$ & $\surd$ & $\times$ & $\times$ & $\times$ \\[0.4 mm]\hline\hline
\end{tabular}
\end{table}
Here we have assumed that the processes are considered in the gluon dominated regions and the color singlet contributions dominate in case of quarkonium production. 

In contrast to dijet production in DIS, dijet production in $p\,A$ collisions, probes a combination of six unpolarized gluon TMDs \cite{Kotko:2015ura}, which reduce to a combination of DP and WW TMDs in the large-$N_c$ limit. It will be extremely hard to extract these six TMDs from experiment. In addition, there is the issue of factorization breaking contributions~\cite{Rogers:2010dm} to worry about for this process, certainly if the $x$ values probed are only moderately small, although at present the magnitude of such contributions is unknown. 

\subsection{Linearly polarized gluons in unpolarized hadrons at small $x$}
The distributions $h_{1\, WW}^{\perp \, g}\equiv h_1^{\perp \, g \, [+,+]}= h_1^{\perp \, g \, [-,-]}$ and $h_{1\, DP}^{\perp \, g} \equiv h_1^{\perp \, g \, [+,-]}= h_1^{\perp \, g \, [-,+]}$ are expected to be unequal as well. In the MV model the DP $h_1^{\perp \, g}$ distribution turns out to be maximal for all transverse momenta, i.e.\ $\bm{k}_T^2 h_{1\,DP}^{\perp \, g}/(2M_p^2) = f_{1\,DP}$, whereas the WW $h_1^{\perp \, g}$ distribution is maximal at large $k_T \gg Q_s$, but suppressed logarithmically in the saturation region ($k_T \ll Q_s$)~\cite{Metz:2011wb}:
\beq
\frac{\bm{k}_T^2}{2M_p^2} \frac{h_{1\, WW}^{\perp \, g}}{f_{1\, WW}} \propto \frac{1}{\ln Q_s^2/\bm{k}_T^2}.
\eeq
In practice, this suppression may be very moderate, except at very small $x$ \cite{Dumitru:2015gaa}. 
 
In the small-$x$ $k_T$-factorization approach the linear gluon polarization is also maximal and moreover positive, leading to \cite{Catani:1990eg}:
\beq
 \Gamma_U^{\,\mu\nu}(x,\bm{k}_T )_{\rm max\ pol}= \frac{k_T^\mu k_T^\nu}{\bm{k}_T^2}\,x\,f_1^{\, g}(x,\bm{k}_T^2).
\eeq
In \cite{Catani:1990eg} both $\gamma^* g^* \to Q \, \overline{Q}$ and $g^* g^* \to Q \, \overline{Q}$ subprocesses were considered, which does not seem to influence the polarization state of the gluons, hence no difference between the WW and DP distributions is obtained or considered in this approach. 

The perturbative large-$k_T$ tail of the WW and DP $h_1^{\perp \, g}$ is the same and shows a $1/x$ growth. This means that the ratio $h_1^{\perp \, g}(x,\bm{k}_T^2)/f_1^{g}(x,\bm{k}_T^2)$ is not suppressed at large $\bm{k}_T^2$ for decreasing $x$. This does not mean however that the observable effects at large transverse momentum are large at small $x$. In TMD factorized expressions Fourier transforms of TMDs enter (cf.\ Eq.\ (\ref{calRQT})) and these are integrals over all transverse momenta, such that effects of linear gluon polarization always suffer from (at least) an $\alpha_s$ suppression w.r.t.\ the unpolarized distribution which has an $\alpha_s^0$ contribution that is a constant in $b$ at fixed scale (for more details cf.\ \cite{Boer:2011kf}). 

In the saturation region the WW and DP distributions descibe expectation values of operators in the Color Glass Condensate (CGC). Both $h_{1\, WW}^{\perp \, g}$ and $h_{1\, DP}^{\perp \, g}$ thus reflect information about the average polarization state of the gluons in the CGC. As there are processes in which the linear polarization becomes maximal as $x \to 0$, the CGC must be maximally polarized, but it depends on the process (and the accompanying operator) whether one actually probes this polarization fully or not. As the MV model indicates, the processes that probe the DP distributions are most sensitive to the linear polarization. In Table \ref{h1table} we list for a selection of processes whether they probe the WW and/or DP $h_1^{\perp \, g}$ distributions. 
\begin{table}[htb]
\centering
\caption{Selection of processes that probe the WW or DP linearly polarized gluon TMD at small $x$.}
\label{h1table}       
\begin{tabular}{|l|c|c|c|c|c|c|}
\hline
\hline
{} & {} & {} & {} &{} & {}\\[-2 mm]
{}& $p p \to \gamma\, \gamma \,X$ & $pA\to \gamma^*\, {\rm jet}\, X$ & $e \, p \to e'\, Q \, \overline{Q} \, X$ & $p p \to \eta_{c,b} \, X$ & $pp \to J/\psi\, \gamma\, X$ \\[0.4 mm]
{} & {} & {} & $e \, p \to e'\, j_1 \, j_2 \, X$ & $p p \to H \, X \hspace{2mm} $ & $pp \to \Upsilon\, \gamma\, X\hspace{3mm} $ \\[0.4mm]\hline
$h_1^{\perp\,g \, [+,+]}$ (WW) & $\surd$ &$\times$ & $\surd$ & $\surd$ & $\surd$ \\[0.6 mm]\hline
$ h_1^{\perp\,g\, [+,-]}$ (DP) & $\times$ &$\surd$ & $\times$ & $\times$ & $\times$ \\[0.4 mm]\hline\hline
\end{tabular}
\end{table}

The processes of DIS, DY, SIDIS, light hadron production and $\gamma+{\rm jet}$ production in $pp$ or $pA$ collisions do not probe $h_1^{\perp \, g}$ in leading power \cite{Boer:2009nc}, but $\gamma^*+{\rm jet}$ production in $pp$ or $pA$ collisions does in the kinematic regime where gluons in one proton dominate~\cite{Metz:2011wb}, as do $J/\psi+\gamma$ and $\Upsilon + \gamma$, thanks to the additional scale present ($Q$, $M_{J/\psi}$ and $M_\Upsilon$, respectively). The gluon rescatterings for the subprocess $q \, g \to \gamma^* \, q$ lead to a gluon correlator with a $+$ and $-$ link, hence to the DP gluon TMD. On the other hand, $g \, g \to J/\psi\, \gamma$, $g \, g \to \Upsilon\, \gamma$, $\gamma^* g \to Q \, \overline{Q}$ and  $\gamma^* g \to q \, \bar{q}$ all come with the WW gluon TMD. Since there are different expectations for the WW and DP $h_1^{\perp \, g}$ inside and outside the saturation region, it would be very interesting to compare these distributions as a function of transverse momentum and $x$. 

Expressions for dijet and heavy quark pair production in DIS at a high-energy EIC including $h_1^{\perp \, g}$ contributions can be found in \cite{Pisano:2013cya,Boer:2016fqd} for general $x$ and in \cite{Metz:2011wb,Dumitru:2015gaa} for small $x$. 
As for the unpolarized distributions dijet and heavy quark pair production in $p\,p$ and $p\,A$ collisions probe a combination of six distinct $h_1^{\perp \, g}$ distributions and have the problem of factorization breaking contributions. Similarly, the process $gg \to [\overline{Q}Q] g$, relevant for quarkonium plus jet production in $p\,p$ and $p\,A$ collisions, has a more intricate link structure that has not been investigated yet and it is unclear whether color singlet quarkonium production dominates, which is necessary to avoid factorization breaking color entanglement. Finally, we mention that in \cite{Mukherjee:2015smo} the processes of $p p \to J/\psi \, X$ and $p p \to \Upsilon \, X$ were considered in the Color Evaporation Model (CEM), as promising probes of the linear gluon polarization at RHIC, LHCb, and AFTER@LHC. 

\subsection{Gluon Sivers effect at small $x$}
Unlike the unpolarized and linearly polarized gluon distributions, the Sivers function does not remain invariant under interchanging $+$ and $-$ links (i.e.\ under a $P$ and $T$ transformation). Rather for the Sivers function (and the other T-odd TMDs), it holds that  $f_{1T}^{\perp \, g \, [+,+]}= - f_{1T}^{\perp \, g \, [-,-]}$ and $f_{1T}^{\perp \, g \, [+,-]}= - f_{1T}^{\perp \, g \, [-,+]}$, as we have discussed before. In Table \ref{f1Ttable} we list for a selection of processes which gluon Sivers TMD they probe.
\begin{table}[htb]
\centering
\caption{Selection of processes that probe the WW or DP Sivers gluon TMD at small $x$.}
\label{f1Ttable}       
\begin{tabular}{|l|c|c|c|c|c|c|c|c|}
\hline
\hline
{} & {} & {} & {} &{} & {} & {} \\[-2 mm]
{} & DY & SIDIS & $p^\uparrow\, A \to h\, X$ & $p^\uparrow A\to \gamma^{(*)}\, {\rm jet}\, X$ & $p^\uparrow p \to \gamma\, \gamma \,X\hspace{7.5mm}$ & $e \, p^\uparrow \to e'\, Q \, \overline{Q} \, X$  \\[0.4 mm]
{} & {} & {} & {} & {}  & $p^\uparrow p \to J/\psi \, \gamma \,X\hspace{3.5mm}$ & $e \, p^\uparrow \to e'\, j_1 \, j_2 \, X$  \\[0.4mm]
{} & {} & {} & {} & {}  & $p^\uparrow p \to J/\psi \, J/\psi \,X$ & {}  \\[0.5mm]\hline
$f_{1T}^{\perp\,g \, [+,+]}$ (WW) & $\times$ & $\times$ & $\times$ & $\times$ & $\surd$ & $\surd$ \\[0.6 mm]\hline
$ f_{1T}^{\perp\,g\, [+,-]}$ (DP) & $\surd$ & $\surd$ & $\surd$ & $\surd$ & $\times$ &$\times$ \\[0.4 mm]\hline\hline
\end{tabular}
\end{table}

There can be no single transverse spin asymmetry in DIS, therefore, it is not sensitive to the Sivers effect either. On the other hand, semi-inclusive DIS is sensitive to the Sivers effect. 
At an EIC the golden channel for the gluon Sivers effect is dijet or heavy quark pair production in DIS, which involves the WW gluon Sivers TMD. 
Based on the behavior of the large-$k_T$ tail of the TMDs, it appears \cite{Boer:2015pni} that in the small-$x$ limit the WW gluon Sivers TMD becomes suppressed by a factor of $x$ w.r.t.\ the unpolarized gluon TMD, while the DP gluon Sivers TMD grows as fast as the unpolarized gluon one. The DP gluon correlator involves a $+$ and a $-$ link, thus enclosing a rectangular area on the lightfront. In \cite{Boer:2015pni} and \cite{Boer:2016xqr} it is shown that as $x\to 0$ the DP gluon TMD correlator becomes proportional to a correlator of a pure Wilson loop $U^{[\Box]}= U^{[+]}_{[0,\xi]} U^{[-]}_{[\xi,0]}$, which measures the flux of the chromo-electromagnetic field through the rectangular loop. In a sense the single transverse spin asymmetry measurement at small $x$ is the QCD analogue of the measurement of the Aharonov-Bohm effect, where the effect of a nonintegrable QED phase factor is probed. This makes the single spin asymmetry at small $x$ a rather fundamental observable in QCD. 

The $C$-odd combination $ U^{[\Box]} - U^{[\Box]\dag}$ corresponds to the odderon operator. In the leading twist TMD formalism considered here it only contributes for transversely polarized protons in the limit $x\to 0$ \cite{Boer:2016xqr}:
\beq
\Gamma_{(d)}^{(T-{\rm odd})} \equiv
\left(\Gamma^{[+,-]}-\Gamma^{[-,+]}\right) \propto {\rm F.T.}\,
 \langle P, S_T |  {\rm Tr} \left [
U^{[\Box]}(0_T,y_T) - U^{[\Box] \dag}(0_T,y_T) \right ]
 |P, S_T \rangle.
\eeq
The absence of a spin-independent odderon in this approach simply indicates that it is suppressed w.r.t.\ the unpolarized gluon distribution in the small-$x$ limit at leading twist. 

The spin-dependent odderon correlator gives rise to single transverse spin asymmetries in $p^\uparrow p $ and $p^\uparrow A$ 
scattering in the small-$x$ regime, e.g.\ in backward hadron production. Since the odderon is $C$-odd, 
for $gg$-dominated processes (i.e.\ for not too large transverse momentum) one should consider final states that are not $C$-even, such as charged hadron production. 
$C$-odd quarkonia may also offer suitable probes. $C$-odd quarkonia are produced in a color octet state from two gluons (in which case also $J=1$ is allowed), but can be in a color singlet state when produced from three gluons. At small-$x$ the gluon density is so high that this additional gluon may not be a concern. 

Single spin asymmetries in $e \, p^\uparrow \to e'\, J/\psi \, X$ have been investigated in the CEM in \cite{Godbole:2013bca} and in the Color Octet Model in \cite{Mukherjee:2016qxa}, yielding similar results at leading order. 

Here we have limited the discussion to TMD factorizing processes or to processes for which hydrid factorization descriptions have been considered in the small-$x$ limit. Other approaches to describe the single transverse spin asymmetries have also been applied, e.g.\   
for $\eta_{c,b}$ production using the twist-3 approach \cite{Schafer:2013wca} or for $D$-meson production using the so-called generalized parton model approach \cite{Godbole:2016tvq}.

\section{Summary}
Quarkonium production offers good possibilities to study various aspects of gluon TMDs. It can be used to probe the unpolarized gluon TMD $G^{(1)} \equiv f_1^{\,g \, [+,+]}$, which has not yet been extracted from experiments. $C$-even bottomonium production ($\chi_{b0/2}$ and $\eta_b$) at LHC allows to study the effects of linearly polarized gluons rather cleanly by means of ratios of ratios, for which hadronic uncertainties cancel. There is no need to measure angular distributions, nor does one need to worry about TMD evolution. $C$-even bottomonium production at LHC also allows a study of the size of factorization breaking contributions in case of $p$-wave states. Experimentally the measurement of $\chi_{b0}$ and $\eta_b$ production is of course very challenging. The TMDs accessed in this way are of the WW type (${\scriptstyle [+,+]}$), which allows direct comparison to certain $\cos 2\phi$ asymmetries in heavy quark pair and dijet production in DIS at a high-energy EIC, which probe the same functions. At small $x$ a significant change in behavior around $k_\perp \sim Q_s$ is expected for $h_{1\, WW}^{\perp \, g}$. Processes that probe the DP-type functions are expected to be more relevant at small $x$ and to exhibit larger effects from the polarization of the CGC. 

In case of transversely polarized hadrons, single spin asymmetries allow to probe the Sivers effect. $J/\psi$ pair production at AFTER@LHC can be used to access the WW gluon Sivers TMD, as an alternative to using photons. It allows to test a sign-change relation w.r.t.\ the WW gluon Sivers TMD probed at an EIC in open heavy quark pair production. The WW gluon Sivers TMD is expected to be suppressed by a factor of $x$ in the small-$x$ limit, although expressions that are valid for the entire $k_T$ range are still lacking. The DP gluon Sivers TMD does not suffer from this suppression. In the limit of small $x$ it is related to the spin-dependent odderon, which is given by a single Wilson loop correlator. This leads to a rather striking simplification of the theoretical description of single spin asymmetries at small $x$. Given the $C$-odd nature of the odderon, $C$-odd quarkonia are likely suited as probes, although this requires further study.    

\begin{acknowledgement}
I would like to thank Jean-Philippe Lansberg, Cristian Pisano, and Jian Zhou, for fruitful discussions and/or collaboration on several of the topics presented here. 
\end{acknowledgement}

\end{document}